\documentstyle[twoside,fleqn,espcrc2,epsf]{article}

\newcommand{\ewxy}[2]{\setlength{\epsfxsize}{#2}\epsfbox[10 60 640 570]{#1}}

\newcommand{\be}{\begin{equation}}
\newcommand{\ee}{\end{equation}}

\newcommand{\lqcd}{{\Lambda_{QCD}}}

\newcommand{\nl}{\nonumber \\}

\newcommand{\csw}{{C_{SW}}}

\newcommand{\muoW}{{\mu_1^W}}
\newcommand{\mutW}{{\mu_2^W}}
\newcommand{\mutC}{{\mu_2^C}}

\newcommand{\Ord}{{\cal O}}

\newcommand{\mev}{{\rm MeV}}
\newcommand{\gev}{{\rm GeV}}

\newcommand{\asig}{{a\sqrt{\sigma}}}

\newcommand{\AmS}{{\protect\the\textfont2
A\kern-.1667em\lower.5ex\hbox{M}\kern-.125emS}}

\title{
SCRI Results With the Tadpole-Improved Clover Action
}

\author{
Sara~Collins\address{Dept. of Physics, Glasgow University, Glasgow G12 8QQ, 
Scotland},
Robert~G.~Edwards${}^{\rm b}$,
Urs~M.~Heller\address{SCRI, Florida State University,
Tallahassee, FL 32306-4052, USA},
and John~Sloan\address{Dept. of Physics and Astronomy, University of Kentucky,
Lexington, KY 40506-0055, USA}${}^{,}$\thanks{Speaker at the Workshop.}
}

\begin{document}

\begin{abstract}

I discuss a study performed by the SCRI lattice gauge theory group
which compared light hadron spectroscopy using the Wilson and Clover 
fermionic actions.  We showed that a Clover coefficient chosen
using tadpole-improved tree-level perturbation theory effectively
eliminates the $\Ord(a)$ discretization errors present in the Wilson
action.  We found that discretization errors in light spectroscopy for
both the Wilson and Clover actions are characterized by an energy
scale $\mu$ of about $200$-$300$ $\mev$, indicating that these
errors can be reduced to the $5$\% level by using the Clover action
at an inverse lattice spacing of about $1.3$ $\gev$.

\end{abstract}

\maketitle

\section{INTRODUCTION}
In this talk, I will discuss the results of an ongoing project at SCRI
to understand the systematic errors in lattice light spectroscopy 
calculations.

The short term goal of our program has been to determine the value of 
the lattice spacing at which the Clover lattice fermion action
reduces discretization errors to the 5\% level.  Once this lattice spacing 
is known, we can perform dynamical simulations on a large physical volume
and attempt to match our results to experiment.

To determine this lattice spacing, we have concentrated on two sub-projects.
First, we have performed a detailed comparison of light spectroscopy
using the Wilson and Clover discretizations on a pre-existing gluonic ensemble,
generated by the HEMCGC collaboration, which has an inverse lattice spacing
of about $2$ $\gev$ and includes the effects of two flavors of (staggered)
light quarks.  The purpose of this study was to see if, at typical lattice
spacings, the Clover discretization significantly improved the
discrepancies with experiment found in Wilson simulations, particularly
in ratios of observables which probe different energy scales (such as the
ratio of bottomonium splittings to the proton mass).  Second, we have
used an improved gluonic action on coarse lattices to determine the
size of the scaling violations directly where they are large.

This talk is organized as follows.  In section 2, I discuss
technical details of our calculations.
In section 3, I discuss the techniques we used in the scaling analysis
of our quenched spectroscopy data, while in
section 4, I present the results of these calculations.
In section 5, I present the results
of our comparison of Wilson and Clover spectroscopy, including the effects of
two flavors of dynamical fermions, at an intermediate lattice spacing.
In section 6, I summarize our conclusions.

\section{THE SIMULATIONS AND FITTING PROCEDURES}

We used two different types of gluonic ensemble in this study.  To
determine the size of discretization effects, we generated quenched
ensembles using a one loop $\Ord(a^2)$ improved gluonic 
action~\cite{Luscher_85,Alford_95} at 
six values of $\beta$.  The inverse lattice spacing of these ensembles 
ranged from $1.3$ to $0.5$ $\gev$, corresponding to lattice spacings between
$.15$ and $0.4$ fermi (see table~\ref{tab:quenched_betas}).  The ensemble 
at each $\beta$ consisted of 100 configurations of size 
$16^3\times 32$.
For the dynamical runs, we used $16^3\times 32$ gauge configurations 
generated by the 
HEMCGC collaboration~\cite{hemcgc-ens}, with $\beta=5.6$ and $2$ 
flavors of staggered fermions at a mass of $am=0.01$. The inverse lattice 
spacing for this ensemble has previously been determined to be roughly 
$2$ $\gev$.

\begin{table}[thb]
\caption{Values of $\Ord(a^2)$ improved $\beta$ and corresponding
inverse lattice spacings for quenched study.}
\label{tab:quenched_betas}
\begin{center}
\begin{tabular}{|c|c|c|c|c|c|c|}
\hline
$\beta$        & 7.9 & 7.75 & 7.6 & 7.4 & 7.1 & 6.8  \\
\hline
$a^{-1}(\gev)$ & 1.3 & 1.1  & .96 & .78 & .59 & .48  \\
\hline
\end{tabular}
\end{center}
\end{table}

On all ensembles, both quenched and dynamical, we calculated light 
hadron correlation functions using both Wilson and tadpole improved
Clover quark propagators.
In all cases we used methods involving multi-state fits to 
multiple correlation functions to extract hadron masses, the technical 
details of which are discussed in~\cite{sloan_95}.  

The main focus of this work, especially of the quenched portion, was to 
determine the size of scaling violation effects rather than to 
attempt to reproduce or predict experimental results.  Since discretization
errors should get worse with increasing quark mass (because the typical
energy scales should grow), we did not devote much effort to chiral
extrapolations.  Most of the quenched results I will present here used a 
lowest order (quadratic in $M_{PS}$) chiral fit ansatz, while the dynamical
results were analyzed using both quadratic and cubic ans\"atze.  Gottlieb 
has already presented a detailed discussion of this extremely
important (and ill-controlled) source of systematic errors in his 
talk~\cite{gottlieb_tsu97}, so I will not discuss chiral extrapolations here,
other than to say that our results agree qualitatively with his.

\subsection{Low Quark Masses on Coarse Lattices}
I would like to make two technical comments concerning coarse lattices
before proceeding to the bulk of the talk.  First, for a particular
{\it dimensionful} value of $M_{PS}$, going to a coarse lattice does
not seem to improve the speed of convergence of the fermion inversion
algorithm.  In other words, the condition number of the fermion kernel
seems to be controlled by $m_q/\lqcd$, the quark mass over the QCD
scale, rather than $am_q$, the dimensionless quark mass.  This is a very
reasonable behavior, but it means that the gains to be found in going
to a coarse lattice might be smaller than one might think.

The other comment involves exceptional configurations.  We had originally
planned to perform our quenched coarse lattice runs using $8^3\times 16$
configurations.  At low values of $\beta$, however, we found many exceptional 
configurations. This forced us to increase the lattice size
to $16^3\times 32$, which suppressed the effect.  
This method of increasing the volume to reduce exceptional configurations,
can, however, result in an unacceptable 
increase in computational cost (which scales linearly with the lattice volume
in the regime where finite volume effects are unimportant) unless one is
careful.  This is
because each inversion of a localized hadronic source will only 
sample a roughly $(1{\rm fm})^3$ volume of the lattice, so there is no
gain in statistics when one increases the lattice volume.  The solution
is to smear the hadron around multiple origins (separated by some multiple
of the hadron's radius) on the initial timeslice.  Naively this would 
increase the computational cost by the number of origins used, but one
can invert the quark propagators from all origins simultaneously by 
superposing them with a random $Z(3)$ phase~\cite{z3noise}.  When using this
method the computational cost of inverting from multiple origins is
the same as the cost to invert from a single origin.  

In practice, this means that one can double the spatial extent of 
the lattice (decreasing both
finite volume and exceptional configuration effects) {\it without
increasing the computational cost of achieving equal statistics}!  This
is because, although each inversion will be eight times as expensive,
it will also yield eight times as many statistically independent measurements.
This means that (ignoring memory limitations, thermalization times,
and the reduction of HMC acceptance rates on larger volumes) one can 
{\it always} effectively eliminate finite volume 
effects from one's simulations at no additional CPU cost by increasing 
the lattice volume and using
the superposition trick.  Although the superposition trick is vital on
coarse lattices (where the lattice volume must be large), it should also
be used to eliminate finite volume effects on fine lattices.

\section{SCALING FITS}

The main idea in performing extrapolations to the continuum limit,
as well as in attempting to reduce discretization errors through
improvement programs, is to Taylor expand dimensionless physical
ratios around $a=0$.  For example,
\begin{eqnarray}
\lefteqn{{{M_{\rho}}\over{\sqrt{\sigma}}}(\asig) }      \nl
&
\!\!\!\!\!\!\!\!
=&
\!\!\!\!\!\!\!\!
M(0)\left[1 + {{M'(0)}\over{M(0)}}(\asig) 
+ {{M''(0)}\over{M(0)}}(\asig)^2 + \dots\right]         \nl
&
\!\!\!\!\!\!\!\!
=&
\!\!\!\!\!\!\!\!
M(0)\left[1 + C^{(1)}a + C^{(2)}a^2 + \dots\right],
\label{eqn:taylor}
\end{eqnarray}
where $M(0)$, $M'(0)$, $\dots$ are the continuum limits of the ratio 
and its derivatives (note that I will work in units of the string tension 
$\sigma$ when discussing scaling, even though it is not an experimental
observable).
Continuum extrapolations attempt to truncate the expansion at some order
(usually first) and fit to data obtained at different $a$'s to determine M(0).
Improvement programs modify the action so as to reduce or eliminate the
low order coefficients such as $C^{(1)}$ and $C^{(2)}$.  For the 
(non-perturbatively tuned) Clover action, the leading discretization error
is $C^{(2)}a^2$.  Our strategy in this project has been to determine 
$C^{(2)}$ by running on coarse lattices with large $a$.  Once we have
determined $C^{(2)}$, we can choose a value of $a$ for our production
runs at which the discretization errors are comparable to the other
errors in our simulation.  In particular, it is a waste of computational
effort to run at values of $a$ where the attainable statistical errors
are an order of magnitude larger than the discretization errors; higher
statistics can be obtained for the same effort on coarser lattices, while
the discretization errors will still be negligible.

\subsection{$\mu$}
Discretization errors in lattice calculations are caused by irrelevant,
higher dimension operators appearing in the cut-off action.  In general,
physical observables will have some characteristic infra-red energy scale,
$\mu$, which is the typical energy scale of the quantum fluctuations
important to the observable.  For light hadronic observables, we expect this
scale to be set by $\lqcd$ or the constituent quark mass, i.e. $\mu$ should
be several hundred $\mev$.  

If we rewrite the dimensionful coefficients of equation~\ref{eqn:taylor},
$C$, in terms of dimensionless coefficients $D$,
i.e.~$C^{(n)} = D^{(n)}\mu^n$, then a reasonable discretization should
have $D$'s which are $\Ord(1)$.  If we assume that this is the case 
(i.e.\ set $D=1$), then each coefficient in equation~\ref{eqn:taylor} gives
us an independent estimate of $\mu$:
\be
\mu_n \equiv |C^{(n)}|^{1/n}.
\label{eqn:mudef}
\ee
If the different estimates $\mu_n$ are in rough agreement, then it
is reasonable to estimate the higher order $\Ord(a^N)$ corrections
to be about $(\mu_n a)^N$.  Conversely, if the estimates $\mu_n$ wildly
disagree, then we know that the coefficients of the $a$ expansion
are not very well behaved, and we should be very cautious in inferring
the size of higher order (unmeasured) discretization errors.  In our
quenched study, we found that $\mu_n$'s for the $\rho$, the nucleon,
and the $\Delta$ were all in the region $200$-$300$ $\mev$, for both
the Wilson and Clover discretizations.  This means that one can expect
discretization errors of about $1$-$2$\% when using the Clover action
at a lattice spacing of $.1$ fm.  If one wants discretization errors
to be about $5$\%, as is reasonable for a first effort at reproducing the
physical spectrum using dynamical fermions, then a lattice spacing
of about $.2$ fm appears to be the best choice.

Note that $\mu$ is closely related to the perturbative $q^*$ of
Lepage and Mackenzie~\cite{PeterPaulPert}.  In particular, one might
consider equation~\ref{eqn:mudef}, for a particular $n$, to be a 
non-perturbative 
definition of $q^*$.  It is not necessary, however, that the physical 
scale $\mu$ be close to our estimates $\mu_n$; a particularly bad
discretization of the action, for example, would yield discretization
errors much worse than the expected $(\mu a)^n$, leading to spuriously
large values of $\mu_n$.  For example, one might expect the $\mu_n$'s
in calculations using Staggered fermions to be about twice those
of Wilson or Clover calculations because of the two hop nature of the
Staggered derivative operator.  It is likely, however, that $\mu$ is 
a (rough) lower 
bound on $\mu_n$, since it is hard to imagine an action with spuriously 
small values of the $D$'s.

\subsection{Scaling Fit Ans\"atze}
We used linear and quadratic ans\"atze for our scaling fits of Wilson
and Clover data.  We called these ans\"atze 1, 2, and 12, corresponding 
to pure
linear (1), pure quadratic (2), or mixed linear plus quadratic (12).
All of our scaling fits are done in units of the string tension;
for physical predictions we must take ratios of extrapolated
values (to remove the string tension, which is not experimentally
measurable).

The first question we attempted to answer was whether the Clover action
with a tadpole improved tree level coefficient is effective in removing
the linear discretization errors of Wilson fermions.  
If the Clover coefficient, $\csw$ is non-perturbatively tuned, then the lowest
order discretization errors will be quadratic.  In our calculations,
however, we used tadpole-improved tree level perturbation theory
to set $\csw$, so we expect $\Ord(\alpha(a)a)$ errors in addition to
$\Ord(a^2)$.  Since $\alpha(a)$ runs fairly slowly, we treated it as
a constant over the range of lattice spacings in our calculation, i.e.~we 
included it in our fit ans\"atze as a linear term with a coefficient which
we expect to be small ($\Ord(\alpha)$).

\begin{table}[thb]
\caption{$Q$ values for linear, quadratic and mixed scaling fits to 
Wilson and Clover vector meson masses.   $V$ indicates the
benchmark vector particle at $M_{PS}/M_V = 0.7$.}
\label{tab:ansatz_Q}
\begin{center}
\begin{tabular}{|c|c|c|c|c|}
\hline
State & Action & linear & quadratic & mixed \\
\hline
$\rho$ & Wilson & .41 & $\approx 10^{-6}$  & .34 \\
\cline{2-5}
       & Clover & .20 & .98        & .94 \\
\hline
$V$    & Wilson & .28 & $\approx 10^{-10}$ & .45  \\
\cline{2-5}
           & Clover & .04 & .94        & .94  \\
\hline
\end{tabular}
\end{center}
\end{table}

Table~\ref{tab:ansatz_Q} contains the values of the confidence level,
$Q$, for scaling fits of
the vector mass at all six $\beta$ values, using the 1, 2, and 12 ans\"atze.
Both Wilson and Clover results are shown for vector masses corresponding to 
the $\rho$
(i.e.~$M_{PS}/M_V = M_\pi/M_\rho$) and to a benchmark value agreed to
at the recent Seattle workshop~\cite{seattle}, for which chiral 
extrapolations are not
required ($M_{PS}/M_V = 0.7$).  For the Wilson fits it is clear, as expected,
that a linear term is necessary; the pure quadratic ansatz is clearly
ruled out by the small $Q$ values.  For the Clover fits, however, one cannot
draw a definitive conclusion from these $Q$ values alone; both the linear
and quadratic fits to the $\rho$ have acceptable $Q$, although the linear
$Q$'s (especially $.04$ for the $V$) are low.  The message to be drawn from 
Table~\ref{tab:ansatz_Q}
is that it is {\it extremely} difficult to distinguish between analytic
functional forms to be used in extrapolations by looking at confidence 
levels for fits to a single curve; roughly 2 Gflops$\cdot$yr of CPU time 
were required to generate 
the Clover data yet these fits still are unable to distinguish linear from 
quadratic behavior.  This observation is important not only when doing
lattice spacing extrapolations, but also when doing chiral extrapolations.
I will show later that simultaneous fits to both Wilson and Clover data
do rule out the pure linear Clover ansatz; unfortunately we do not
know how to apply this technique to chiral extrapolations.

\begin{figure}[htb]
\centerline{\ewxy{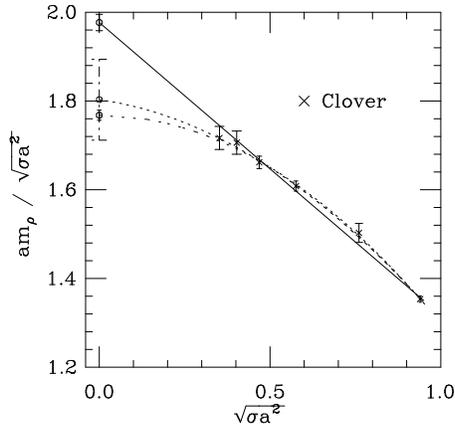}{80mm}}
\caption{
Results of different scaling fits to the $\rho$ mass.
}
\label{fig:mult_ansatz}
\end{figure}

Although the linear and quadratic ans\"atze cannot be distinguished by $Q$,
they result in significantly different continuum extrapolations.  
Figure~\ref{fig:mult_ansatz} shows the results of the three different
fits to the Clover $M_\rho$ data, along with the continuum extrapolations.
It is obvious that the linear and quadratic extrapolations differ by
many standard devations.  This is a serious problem, since it shows 
that there is
a systematic error associated with choice of fit ansatz which can be
many (statistical) $\sigma$.  

Our analytic knowledge tells us that the (perturbatively tuned) Clover
action should reduce but not eliminate the $\Ord(a)$ errors.  The mixed
ansatz permits both the large quadratric and suppressed linear terms to
appear in the extrapolation.  As can be seen from 
figure~\ref{fig:mult_ansatz}, the linear coefficient is consistent with
zero but the extrapolated value has a much larger statistical uncertainty.
This is because (as in all other cases in table~\ref{tab:ansatz_Q}) the 
mixed fit 
is underdetermined; an ansatz with fewer parameters (linear for Wilson 
or quadratic for Clover) fits the same data with good $Q$.  We expect the
true uncertainty of the Clover extrapolation to fall somewhere between
the quadratic and mixed error bars.  In presenting extrapolated values I
will quote both estimates of the uncertainty.

\subsection{Joint Fits}

Until now, I have presented the Clover and Wilson results separately.
The zeroth order coefficient in both cases, however, is 
just the continuum limit of the quantity being fit.  This value should
not depend on the discretization scheme used,
so we can obtain more constrained fits by simultaneously
fitting Clover and Wilson data:
\begin{eqnarray}
M_W(\asig) &=& M(0)\left[1 + (\mu_1^Wa) + (\mu_2^Wa)^2 \right] \nl
M_C(\asig) &=& M(0)\left[1 \hphantom{+ (\mu_1^Wa)} + (\mu_2^Ca)^2 \right],
\label{eqn:joint_ansatz}
\end{eqnarray}
where $M$ is the quantity being fit (i.e.~$aM_V/\asig$) and the
subscript indicates data from Wilson (W) or Clover (C) discretizations.
In this example, I have used the 2 ansatz for the Clover data, and
the 12 ansatz for Wilson.  Although the Wilson and Clover data
at a particular value of $a$ are correlated, we have not included these
correlations in our joint fits.  This could lead us, in principle, to 
overestimate the confidence levels of our fits, but we expect this 
effect to be small.

\begin{table}[thb]
\caption{$Q$ values for joint scaling fits to Wilson and Clover
vector meson masses.   $V$ indicates the
benchmark vector particle at $M_{PS}/M_V = 0.7$.}
\label{tab:joint_Q}
\begin{center}
\begin{tabular}{|c|c|c|c|c|}
\hline
State      &\multicolumn{4}{|c|}{Ansatz} \\
\cline{2-5}
           & Wilson &\multicolumn{3}{|c|}{Clover} \\
\cline{3-5}
           &        & linear             & quadratic          & mixed \\
\hline
$\rho$     & linear & $\approx 10^{-26}$ & .016               & .69 \\ 
\cline{2-5} 
           & mixed  & .002               & .85                & .78 \\
\hline
$V$        & linear & $\approx 10^{-72}$ & $\approx 10^{-12}$ & .16  \\
\cline{2-5}
           & mixed  & .001 & .98         & .96  \\
\hline
\end{tabular}
\end{center}
\end{table}

Table~\ref{tab:joint_Q} presents the $Q$'s for joint fits using all
three ans\"atze for Clover and the linear and mixed ans\"atze for Wilson
data.  The quadratic Wilson ansatz is not considered because it has already
been ruled out in table~\ref{tab:ansatz_Q}.  It is now obvious that the
pure linear ansatz for Clover is also ruled out; its extrapolated value cannot
be made to agree with that of any of the Wilson ans\"atze.  Although the 
12\_1 (mixed Clover, linear Wilson) fits have reasonable Q, they are unstable 
when more parameters are added; the parameters differ from those in
the 12\_12 fits by $1-2$ standard deviations for the $\rho$ and 
$3\sigma$ for the $V$
(using the underdetermined 12\_12 error bars).  As we saw when comparing 
the Clover fits in table~\ref{tab:ansatz_Q}, this is an indication of an 
incorrect ansatz.  This behavior should 
be compared with that of the 2\_12 fits, all of whose parameters 
agree with those in the 12\_12 fits.  
This is entirely consistent with
our analytic expectations; we are at a coarse enough lattice spacing that
we can resolve the quadratic errors in Wilson fermions in addition to
the leading linear errors, which the Clover term suppressed sufficiently
to allow a pure quadratic ansatz.  When an additional linear term is
included (i.e.~12\_12 rather than 2\_12 ansatz), the parameters are 
essentially unchanged, albeit with larger error bars, while the linear
term is consistent with zero.

In presenting the scaling results that follow, I will use central values 
from the 2\_12 fits and quote error bars from both the 2\_12 and 12\_12 
fits as ``best case'' and ``worst case'' uncertainties.

\section{SCALING RESULTS}

\begin{figure}[htb]
\centerline{\ewxy{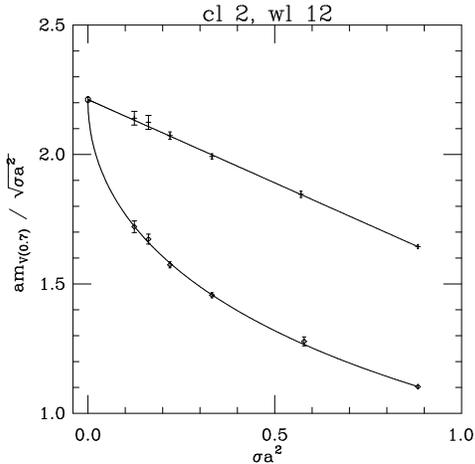}{70mm}}
\caption{
Joint scaling fit to the benchmark vector mass, using a pure
quadratic ansatz for Clover(+) and mixed
ansatz for Wilson($\diamond$) data.
}
\label{fig:joint_vec}
\end{figure}

Figure~\ref{fig:joint_vec} shows the results of a joint 2\_12 fit to
the benchmark vector mass, with the string tension used to set the scale.
The most striking feature is that, at reasonable $\beta$, the Clover 
term has succeeded in eliminating most of the discretization effects,
even though its
coefficient was set using tree-level (tadpole-improved) perturbation
theory.  Note that the two highest $\beta$ values correspond to a 
inverse lattice spacing of about $1$ $\gev$; a $6$ $\gev$ inverse 
lattice spacing would be required before the Wilson discretization errors
are reduced to the same level. The slope parameters from 
the fit, $\mutC$, $\muoW$ and $\mutW$ as defined in 
equation~\ref{eqn:joint_ansatz}, are $240$, $300$, and
$180$ $\mev$, respectively, where we have taken $\sqrt{\sigma}$ to
be $440$ $\mev$.  The fact that all three estimates are
about the same size is an indication that the discretization scale
$\mu$ is roughly $200$-$300\mev$.  

We have performed similar fits to
the $\rho$ mass, as well as to the masses of the nucleon
and the $\Delta$ and the nucleon-$\rho$ mass ratio.  The results of
these fits are presented in table~\ref{tab:all_mu}, and seem to present
a consistent picture.  The $V$ corresponds to a quark mass near
the strange, so one expects the $\mu$'s to be harder than for the $\rho$,
which contains massless quarks.  The $\mu$'s in table~\ref{tab:all_mu} 
of the $V$ are all slightly harder than those of the $\rho$.
Although this difference is about the same size as the uncertainty in
the $\mu$ estimates, the different vector results are 
correlated so the effect is probably significant.  The baryonic $\mu$'s,
in the range $150$-$200\mev$, seem to be slightly smaller than those of
the vectors, which would lead to smaller discretization errors at a
particular lattice spacing.  This is born out by our dynamical fermion
study (discussed below), where we were unable to distinguish between
Clover and Wilson baryonic masses, but were able to see a difference
in the $\rho$ mass.  The $\mu$ estimate coming from the linear term
in the Wilson nucleon-$\rho$ mass ratio at first appears small, but
this is due to the fact that the linear coefficient for each particle
is negative, leading to a large cancellation in the ratio.

\begin{table}[thb]
\caption{
Estimates of $\mu$ from coefficients of joint scaling fits to 
Wilson and Clover results for various observables. Masses are
measured in units of the square root of the string tension, which was 
taken to be $440$ $\mev$ when estimating $\mu$.
$V$ indicates the benchmark vector particle at $M_{PS}/M_V = 0.7$.
}
\label{tab:all_mu}
\begin{center}
\begin{tabular}{|c|r|r|r|}
\hline
Observable & Clover &\multicolumn{2}{c|}{Wilson} \\
\cline{2-4}
           & $\mutC$ & $\muoW$ & $\mutW$ \\
\hline
$M_\rho$      & 220 & 250 & 130  \\
$M_V$         & 240 & 300 & 180  \\
$M_N$         & 160 & 190 & 160    \\
$M_\Delta$    & 190 & 190 &  80    \\
$M_N/M_\rho$  & 180 &  50 & 230  \\
\hline
\end{tabular}
\end{center}
\end{table}

Overall, we conclude that the $\mu$ for light hadron spectroscopy, using
Wilson-type actions, is roughly $200$-$300\mev$.  If we take a worst-case
estimate of $300\mev$ and assume pure quadratic errors for the Clover
action, this implies that lattice spacings of $2.1$ and $1.3$ $\gev$ 
are required to
obtain discretization errors of $2\%$ and $5\%$, respectively.  To
obtain the same errors with the Wilson action, one would need corresponding
lattice spacings of $15$ and $6$ $\gev$, respectively!  

\begin{figure}[htb]
\centerline{\ewxy{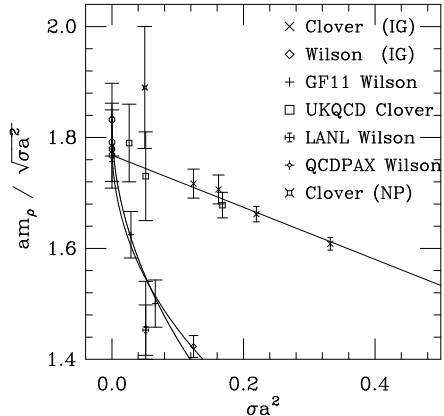}{80mm}}
\caption{
World scaling data for the $\rho$ mass.  IG indicates an improved
gluonic action (our results), while NP uses the chiral ward identity
to tune the Clover parameter.
}
\label{fig:world_rho}
\end{figure}

To illustrate the effectiveness of using the Clover action at 
intermediate lattice spacings, 
figure~\ref{fig:world_rho} shows a sample of world data for 
the $\rho$ mass.  
The discretization errors in our Clover point at $\beta = 7.9$ 
(corresponding to an inverse lattice spacing of about $1.3$ $\gev$)
are about the same size as the statistical errors, and are comparable
to the statistical errors of other calculations.

\begin{table}[thb]
\caption{
Continuum values for quenched light hadron masses and the
nucleon-$\rho$ mass ratio.
Masses are given in units of the string tension.  Statistical errors from 
both 2\_12 and 12\_12 extrapolations are shown.  
$V$ indicates the
benchmark vector particle at $M_{PS}/M_V = 0.7$.
Systematic chiral extrapolation errors for physical masses 
are not estimated (for $M_V$ they should be negligible).  
}
\label{tab:cont_masses}
\begin{center}
\begin{tabular}{|c|l|}
\hline
Observable & Value($\sqrt{\sigma}$) \\
\hline
$M_V$         & 2.122(11)(56)  \\
\hline
$M_\rho$     & 1.758(11)(46)  \\
$M_N$         & 2.30(3)(12)    \\
$M_\Delta$   & 2.93(5)(15)    \\
\hline
$M_N/M_\rho$ & 1.297(17)(81)  \\
\hline
\end{tabular}
\end{center}
\end{table}

Although it was not the main goal of this project, we have obtained the
quenched continuum values of the $\rho$, nucleon, $\Delta$, and $0.7$
benchmark vector masses in units of $\sqrt{\sigma}$, as well as the
nucleon-$\rho$ mass ratio from our joint fits.  These results are presented in 
table~\ref{tab:cont_masses}, with best and worst case error bars shown.  
The benchmark $V$ was chosen to be in the vicinity of computationally
accessible quark masses; it is unlikely to be sensitive to chiral 
extrapolation uncertainties.  The physical hadrons, however, required
a large extrapolation (using an ansatz linear in $M_{PS}^2$); these
results are susceptible to (undetermined) systematic errors arising
from other terms in the chiral expansion.
The only value which can be compared directly to experiment is the 
nucleon-$\rho$ mass ratio.  We
obtain 1.297(17)(81), which is consistent with the experimental value
of 1.22 if one uses the worst case uncertainty.

\section{DYNAMICAL RESULTS}

Our dynamical runs were intended to investigate whether Clover results
on a fine lattice with dynamical fermions (partially) included would 
resolve discrepancies with the experimental spectrum found in
Wilson studies, and to check how large the differences between Wilson
and Clover spectroscopy are on a fine lattice.  The most important of 
these discrepancies is the
disagreement between the lattice spacings obtained from light spectroscopy
and those obtained from $\Upsilon$ splittings~\cite{nrqcd_collab}.  
Previous Wilson calculations on this ensemble~\cite{HEMCGC_wilson} found
inverse lattice spacings of about $2$ $\gev$. Spin-independent bottomonium
splittings calculated on the same ensemble using the NRQCD formalism, however, 
obtained values near $2.4$ $\gev$ - a 20\% discrepancy.  It is unlikely
that this would change if three flavors of dynamical fermion were used,
since quenched results at a similar lattice spacing show roughly the same 
discrepancy.

\begin{figure}[htb]
\centerline{\ewxy{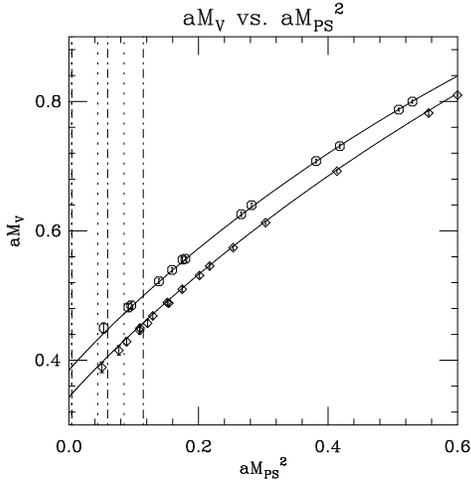}{70mm}}
\caption{
Chiral plot of vector mass on dynamical ensemble comparing 
Wilson($\diamond$) and Clover(o) discretizations
}
\label{fig:rho_mass}
\end{figure}

\begin{figure}[htb]
\centerline{\ewxy{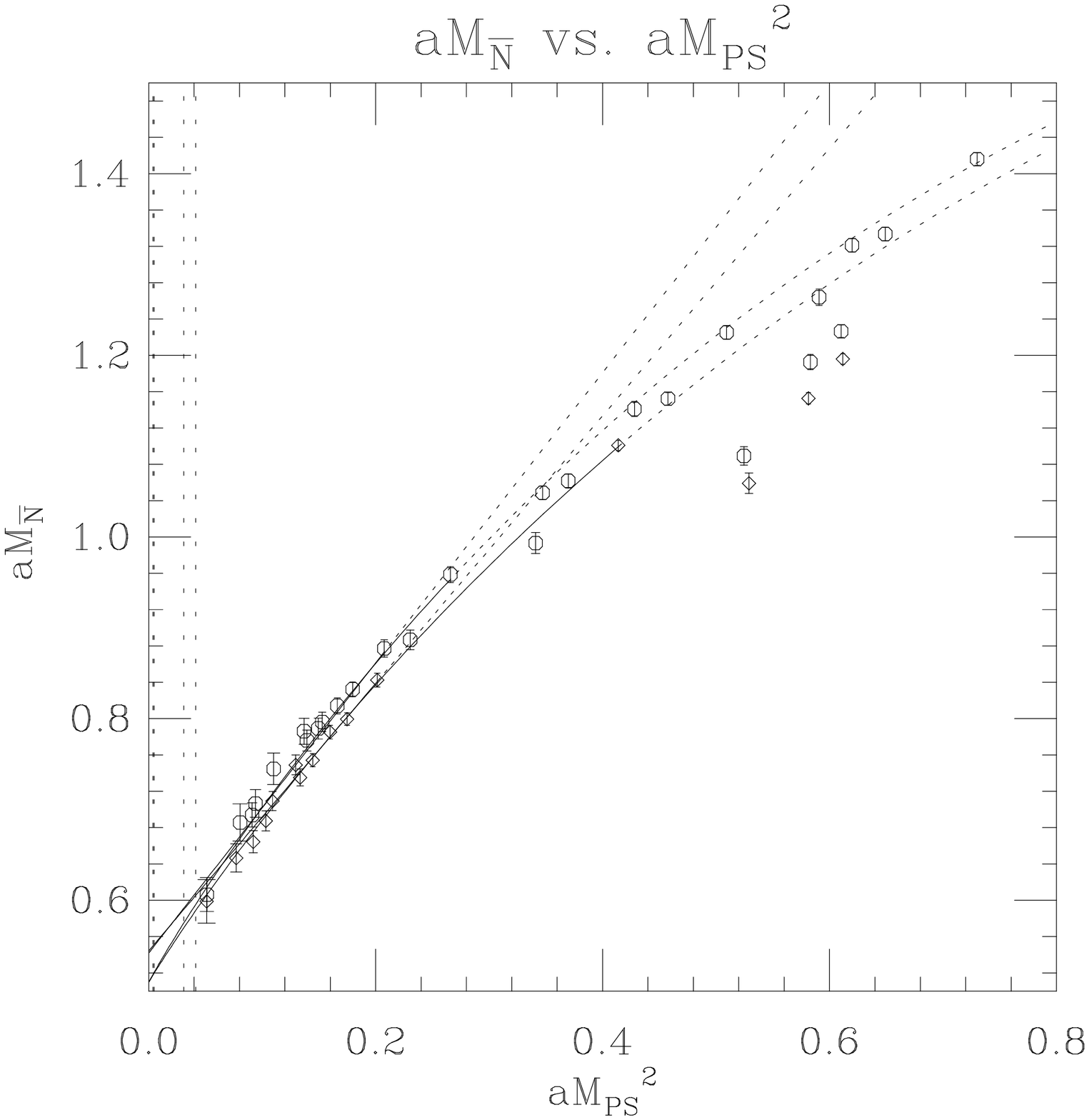}{70mm}}
\caption{
Chiral plot of nucleon mass on dynamical ensemble comparing 
Wilson($\diamond$) and Clover(o) discretizations
}
\label{fig:N_mass}
\end{figure}

\begin{figure}[htb]
\centerline{\ewxy{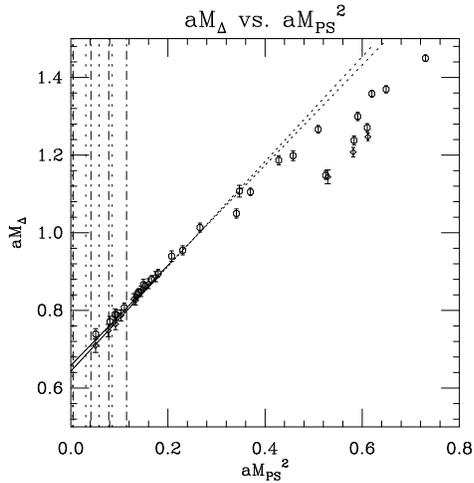}{70mm}}
\caption{
Chiral plot of $\Delta$ mass on dynamical ensemble comparing 
Wilson($\diamond$) and Clover(o) discretizations
}
\label{fig:Delta_mass}
\end{figure}

The good news is that the differences between Wilson and Clover spectroscopy
on the dynamical ensemble are entirely consistent with the $\mu$ value of
about $200$ $\mev$ that we found in the quenched study.  In all observables
we studied, the relative change upon inclusion of the Clover term was
less than or approximately 10\%, i.e.~$(\mu a)$.  In fact, in comparing
chiral intercepts and slopes of the vector, nucleon, and $\Delta$, we
were only able to resolve a difference between the Clover and Wilson
values of the vector intercept, i.e. the mass of the $\rho$.  This
can be seen by examining figures~\ref{fig:rho_mass}, \ref{fig:N_mass}, 
and \ref{fig:Delta_mass}, in which Wilson and Clover chiral plots
of the masses of these three states as a function of pseudoscalar
mass are superposed.  The $\rho$ plot is the only one in which there is a
clear separation between the Clover and Wilson data.
More good news is that we 
found consistency with experiment at the $5$-$10$\% level in the light 
hadron spectrum,  since the Clover $a^{-1}$'s we obtained from different 
light observables ranged from $1.8$-$2.1$ $\gev$.  

The bad news is that these inverse lattice spacings are still inconsistent 
with $\Upsilon$ spectroscopy; the 
20\% discrepancy has not been removed by the Clover discretization.
Since it is not a discretization effect, this discrepancy is probably
due to some other systematic error.  The most likely culprit is finite
volume effects; the spatial extent of our lattice is only 16, which
corresponds to a $1.6$-$1.3$ fm box depending on which lattice spacing 
one uses.  This is much larger than bottomonium, but probably too small
for light hadrons.  Squeezing the light hadronic states would raise 
their masses, resulting in spuriously low inverse lattice spacings.  
Systematic errors in the chiral extrapolation could also be causing
the problem - adding a term linear in $M_\pi$ (which might be induced
by partial quenching effects) to the chiral ansatz
can significantly reduce the extrapolated mass values.  Another possibility
is that the dynamical fermions are not light enough to act
as true up and down quarks.  Whatever the reason for the discrepancy,
higher quality dynamical calculations, especially at larger volumes,
are needed.  It would be a tremendous accomplishment if the community
could demonstrate that lattice QCD can simultaneously reproduce the
spectra of both heavy and light hadron systems.

\section{CONCLUSIONS}

We have shown that a Clover term, with the coefficient chosen using 
tree-level tadpole-improved perturbation theory, eliminates the bulk
of the linear discretization errors in the Wilson action.  At any
value of $\beta$, tadpole-improved Clover calculations will be much
closer to the continuum value than the corresponding Wilson calculation. 

We have shown that the $\mu$'s for light hadron spectroscopy using
the Wilson or Clover actions seem
to be roughly $200$-$300$ $\mev$, with the vector meson discretization
errors slightly larger than those of the baryons.  This means that inverse 
lattice spacings no harder than $1.3$ $\gev$ should be required to reduce 
discretization errors to the 5\% level when using the Clover action.
This determination of $\mu$ can be used by other groups to estimate
the discretization errors they will face at a particular value of
$\beta$ and can also be used to determine the lattice spacings where
further improvements of the Clover action should be attempted.

Finally, we found that it was very difficult to numerically rule out 
linear scaling behavior when looking at Clover data in isolation.  Only
simultaneous fits to both Wilson and Clover data were able to conclusively
rule out a dominant linear behavior in the Clover data in the range of
lattice spacings studied.  These fits indicated that the coefficient of
any sub-dominant $\Ord(a)$ behavior (due to higher order or non-perturbative
corrections to the perturbative Clover coefficient) is small.

\section*{ACKNOWLEDGEMENTS}
This research was supported by DOE contracts DE-FG05-85ER250000
and DE-FG05-96ER40979.
The computer simulations were performed on the CM-2 at SCRI.

\end{document}